\documentstyle[12pt]{article}
\begin{document}
\begin{titlepage}
\begin{flushright}
	USITP-94-02
\end{flushright}
\begin{flushright}
	hep-ph/9402325
\end{flushright}
\vskip .8cm
\begin{center}
{\Large
QCD  corrections  to direct $B\to J/\Psi$ decays}
\vskip .8cm
 {\bf L. Bergstr\"om and P. Ernstr\"om}
\vskip .5cm
Department of Physics\\
Stockholm University\\
Box 6730, S-113 85 Stockholm\\
Sweden
\vskip 1.8cm
\end{center}
\begin{abstract}
We calculate next to leading order QCD corrections to the direct decays
$b\to J/\Psi + X$ and $b\to \eta_c + X$. The strong renormalization
scale dependence is seen to persist also in this order and in fact
the rate is driven to an unphysical negative value at the $M_b$ scale.
We show that this is a consequence of the strong suppression and scale
dependence in the leading order term. Large cancellations
take place between terms from three different orders in $\alpha_s$.
Lacking a third order calculation we are forced to throw away the
leading order term and all but one of the next to leading order terms.
The remaining $c\overline{c}$ gluon emission term could very well be
the dominant one. Even if this is not the case, the picture of
a rate at least as suppressed as in the leading logarithmic
approximation survives.

 \end{abstract}
\end{titlepage}

\newcommand{\beq}{\begin{equation}}
\newcommand{\eeq}{\end{equation}}
\newcommand{\beqa}{\begin{eqnarray}}
\newcommand{\eeqa}{\end{eqnarray}}
\newcommand{\bpsi}{B\to J/\Psi +X}

Recently, there has been renewed interest in the theoretical treatment of
various decays involving $B$ mesons (see e.g.
\cite{braaten,altarelli,buras,bigi}). Since it
now appears likely that dedicated $B$ ``factories'' may be built in the near
future, it is of great importance to match the new level of experimental
accuracy to state-of-the-art calculations within the Standard Model. Only then
can one gain sensitivity to eventual new physics in the study of $B$ decays.
Even within the Standard Model, one has to make sure that the strong
interaction physics involved when relating the Kobayashi-Maskawa origin of
CP violation to measured $B$ decays is well enough understood to establish
this final piece of the six-quark Standard Model.

Presently, most of the $B$ decay phenomenology seems to be reasonably well
understood, with some notable exceptions. First, the measured semileptonic
branching
ratio of $B$ mesons seems to be too low (i.e., the non-leptonic branching ratio
seems to be too high) compared to theoretical calculations. This has already
led the
authors of \cite{bigi} to put the question whether there is new, exotic
physics involved in $B$ decays. Secondly, the QCD-improved effective
non-leptonic
Hamiltonian seems to give a too small a branching ratio of $B\to J/\Psi +X$,
even
if cascade decays are taken into account \cite{braaten}.

Both the semileptonic and nonleptonic decay rates have been calculated in
the next to leading logarithmic approximation. In the calculation of the next
to leading order correction to the  nonleptonic rate, quark masses were put
to zero. After phase space integration this leads to a complete cancellation
of one of three next to leading order terms (each with different Wilson
coefficient structure).
It has been argued that neither higher order QCD
corrections nor  nonperturbative corrections are large enough to explain
the low semileptonic branching ratio \cite{bigi}.

In the case of the  $B\to J/\Psi +X$ decay, a next to leading order calculation
of the perturbative QCD corrections has been lacking (in a previous attempt
\cite{Cox1} the renormalization group was not treated properly \cite{braaten}).

In this paper, we present the results of a calculation of  $B\to J/\Psi +X$
in the next to leading logarithmic approximation (for details, see \cite{Per}).
In a straightforward way we also extend this to $\eta_c$ production.

As will be seen, the reason that the rate for direct $B\to J/\Psi +X$
decay comes out small is that the Wilson coefficient
of the colour singlet effective four fermion operator is strongly
suppressed by the QCD evolution and becomes zero just below the $M_b$ scale.
In the next to leading order, colour octet contributions from $c\overline{c}$
bremsstrahlung appear. These are not suppressed by the QCD evolution
and could therefore be as important as the leading order term.
This is obscured in the next to leading order calculation by
the strong scale dependence of the one-loop contributions, making
the rate negative at the $M_b$ scale. We show that this is a consequence
of the strong suppression and scale dependence of the colour singlet
Wilson coefficient, and propose a slight modification of the resummation
to overcome this problem. After a careful analysis, the next to leading
order calculation is seen to indicate a further suppression of
the rate as compared to the leading order at the $M_b$ scale.
This reinforces the suspicion that the bulk of
$J/\Psi$s produced in $B$ decays originate from other mechanisms.

The nonleptonic effective Hamiltonian describing the $B$ decays we
treat here can be
written \cite{altarelli, buras}

\begin{eqnarray}
{\cal H}_{eff} & = & \frac{G_F}{\sqrt{2}}V_{cb} V^*_{cs}
   \left( C_{1}\left(\mu\right) O_{1} +
          C_{2}\left(\mu\right) O_{2}    \right) \nonumber \\
 & = & \frac{G_F}{\sqrt{2}}V_{cb} V^*_{cs}
   \left( C_{+}\left(\mu\right) O_{+} +
          C_{-}\left(\mu\right) O_{-}    \right) \nonumber \\
 & = & \frac{G_F}{\sqrt{2}}V_{cb} V^*_{cs}
   \left( {1 \over N} C_{0}\left(\mu\right) O_{1} +
                   2  C_{2}\left(\mu\right) O_{8}    \right). \label{Heff}
\end{eqnarray}
Here $V$ are Cabibbo-Kobayashi-Maskawa flavour mixing matrix elements
(we neglect
the Cabibbo suppressed channel $c\to u$ compared to $c\to s$),
$G_F$ is the Fermi weak coupling constant and N is the number of colours
( three for QCD). The O's are four fermion shortdistance operators, all
with the same $(V-A)\otimes(V-A)$ \ $\gamma$ matrix
structure but with different colour structures, while the
$C\left(\mu\right) $'s  are the
corresponding renormalization scale dependent Wilson coefficient functions.
In the last line of Eq.~ (\ref{Heff}) the effective Hamiltonian is written as a
linear
combination of a colour singlet ($O_1$) and a colour octet ($O_8$) operator,
which is the most relevant form for our process.
The operators $O_{\pm}$ are introduced for convenience since they do not
mix under evolution.
We give explicitly

\begin{eqnarray}
  O_{1} & = & (\overline{b}_{\alpha}c_{\beta})_{_{V-A}} \otimes
              (\overline{c}_{\beta}s_{\alpha})_{_{V-A}} \nonumber \\
  O_{2} & = & (\overline{b}_{\alpha}c_{\alpha})_{_{V-A}} \otimes
              (\overline{c}_{\beta}s_{\beta})_{_{V-A}} \nonumber \\
  O_{8} & = & {1 \over 2} O_{2} - {1 \over {2N}} O_{1} \nonumber \\
  O_{\pm} & = & \left( O_{2} \pm O_{1} \right)/2 . \label{Ox}
 \end{eqnarray}
Eqs. (\ref{Heff}) and (\ref{Ox}) lead to the following relations for the
Wilson coefficients:

\begin{eqnarray}
C_{0}\left(\mu\right) & = & {{N+1}\over 2} C_{+}\left( \mu \right) -
                            {{N-1}\over 2} C_{-}\left( \mu \right)  \nonumber
\\
C_{2}\left(\mu\right) & = & \left( C_{+}\left( \mu \right) +
                                   C_{-}\left( \mu \right) \right)/2 .
\label{c0c2}
\end{eqnarray}

In the leading logarithmic approximation, the $C_\pm$ were calculated long
ago
\cite{gaillard,maiani}, with the result

\begin{eqnarray}
L_{\pm}\left( \mu \right) \equiv C_\pm^0 & = &
   \left( \frac{\alpha_{s}\left( M_{W} \right) }
          { \alpha_{s}\left( \mu \right)  } \right)^{d_{\pm}},
\end{eqnarray}
where the anomalous dimensions $d_\pm$ are given by
\beq
d_\pm = { \gamma_\pm^{(0)}  \over 2\beta_0},
\eeq
with
\beq
\gamma_\pm^{(0)} ={\pm 12\left( N\mp 1\right)\over 2 N}
\eeq
and
\beq
\beta_0={11\over 3}N-{2\over 3}n_F,
\eeq
where $n_F$ is the
number of active flavours at the scale $\mu$.

This leads to the decay rate for $B\to J/\Psi + X$ \cite{kuhn}

\beqa
 \Gamma^0_{_{ B \rightarrow J/\psi + X }} & = &
   { L_{0} }^{2} G_{0}, \label{gamma0}
\eeqa
where $L_0$ and $L_2$ are related to  $L_\pm$  in the same way
as $C_0$ and $C_2$ are related to $C_\pm$ (eq. \ref{c0c2}), and with
\beqa
G_{0} & = &  {\cal K}   {{\left( 1 - x \right) }^2}\,\left( 1 + 2\,x \right) ,
\eeqa
where
\beqa
 { \cal K } & = &
    {{{{G_{F}}^2}\,{\left|{V_{cb}}\right|^2}\,
    {\left|{V_{cs}}\right|^2}}\over {96\,{{\pi }^2}}}
    {{{{M_{b}}^3}\,{{R_{s}(0)}^2}}\over {M_{c}}}
    \\
 x & = &   {4 M_c^2 \over M_b^2 }.
\eeqa
Here the nonperturbative coupling of the $c\bar c$ pair to the bound state is
parametrized by
the $S$ wave function at the origin, $R_S(0)$, which can be measured
(including order
$\alpha_s$ corrections) in $J/\Psi$ decays to lepton pairs.

Due to the strong suppression of $L_0$ in the expression for
$\Gamma_0$ (Eq. ~(\ref{gamma0})), the lowest order width is very small, and
strongly dependent on the scale parameter $\mu$ (Fig. 1(a),2). In fact, with
the choice
$\mu\sim 2.3$\, $GeV$, it can even be made to vanish. The hope has been
that the next to leading order corrections would both increase the rate and
make it less sensitive to $\mu$.
 In next to leading order, several new contributions to the Wilson coefficient
functions appear, summarized by the expression
\begin{eqnarray}
 C_{\pm}\left( \mu \right) & = &  L_{\pm} \left( \mu \right)
   \left( 1 + \frac{\alpha_{s}\left( \mu \right)}{4\pi} B_{\pm} \right)
   \left( 1 + \frac{\alpha_{s}\left( M_{W} \right) -
                    \alpha_{s}\left( \mu \right)  } {4\pi} R_{\pm} \right).
\end{eqnarray}

We expect the contribution from penguin induced operators to be small for our
process (for a calculation of their size for the general $\Delta B = 1$
Hamiltonian, see \cite{grinstein,buras2}). Note also that for colour singlet
$c\overline{c}$ production penguin graphs do not appear until second
order in $\alpha_s$.
We work in the 't Hooft-Veltman regularization scheme \cite{HV}, where
$B_\pm$ and $R_\pm$ are related by
\beq
R_\pm=B_\pm+{(N\mp 1)\over 4N\beta_0}\left( -21\pm 57/N\pm
23N\mp 4n_F\right)-(d_\pm /\beta_0)/\beta_1,
\eeq
(see \cite{buras} for a more complete
discussion, including the effects of altering the regularization scheme)
The beta function coefficient $\beta_1$ is given by
\beq
\beta_1={34\over 3 }N^2 - {10\over 2}Nn_f-{(N^2-1)\over N}n_F.
\eeq
We find, in agreement with \cite{buras}
\beq
B_\pm = {\pm B(N\mp 1)\over 2N},
\eeq
where $B=7$ in the 't Hooft-Veltman regularization scheme.
After a calculation of one-loop and bremsstrahlung contributions we find
the rate for $b\to J/\Psi + X$ in the next to leading logarithmic
approximation,
\begin{eqnarray}
 \Gamma_{_{ B \rightarrow J/\psi + X }} & = &
   { L_{0} }^{2} G_{0} +
      {{  \alpha_{s} \left( M_{W} \right) -
            \alpha_{s} \left( \mu \right)      }\over{4\pi}} C_F G_{0} \times
                     \nonumber \\ & &
     \times \left(  2L_{0}L_{2} \left( R_{+} - R_{-} \right) +
                    L_{0}^2 \left( {{2R_{+}}\over{N-1}} + {{2R_{-}}\over{N+1}}
\right) \right)
                     \nonumber \\ & &
+   {{\alpha_{s}( \mu)} \over {4\pi}} C_{F}  \left({ L_{2} }^{2} G_{3}
-   L_{0} L_{2} \left( G_{2} - 2B G_{0}\right) - { L_{0} }^{2} G_{1}
          \right) . \label{full}
\end{eqnarray}
Here  $C_F = (N^2-1)/2N$. $G_3$ stems from real
gluon emission, whereas $G_1$ and $G_2$ are a mixture of real and virtual
contributions. These functions $G_i$, $i=1,2,3$ are given by
\begin{eqnarray}
 G_{1} & = & { \cal K } \biggl(
      2 {{\left( 1 - x \right) }^2}\,\left( 5 + 4\,x \right)  \,\log (1 - x) +
               \biggr. \nonumber \\ & & \biggl.
        \,\, 4x\, \left( 1 - 2\,x \right) \,\left( 1 + x \right)  \,\log (x) +
      \left( 1 - 2\,x \right) \,\left( 1 - x \right) \,\left( 3 + 5\,x \right)
\biggr) +
           \nonumber \\ & &
   G_{0}\,\left( {{4\,{{\pi }^2}} / 3} + 4\,\log (1 - x)\,\log (x) + 8\,{\cal
L}i_{2}(x)
             \right)
    \\
 G_{2} & = & { \cal K } \left(
      {{2 \left( 1 - x \right) \,\left( 34 + 23\,x - 51\,{{x}^2} + 16\,{{x}^3}
\right) }\over
      {2 - x}} +
               \right. \nonumber \\ & &
      {{{{ 32 \left( 1 - x \right) }^3}\,x}\over {2 - x}}\,\log (2) -
      {{{{ 8 \left( 1 - x \right) }^3}\,\left( 3 - {{x}^2} \right) }\over
{{{\left( 2 - x \right) }^2}}}\,
      \log (1 - x) +
           \nonumber \\ & &   \left.
      {{{4{x}^2} \left( 26 - 19\,x + 4\,{{x}^2} \right) }\over {2 - x}}\,
      \log (x) \right) +
       G_{0}\, \left( 12\,\log ( { \mu^2 \over M_b^2 })-4 \right)
   \label{G2} \\
 G_{3} & = &  { \cal K } \left(
   {4\over 9} \left( 1 - x \right) \,\left( 1 + 37\,x - 8\,{{x}^2} \right)
-  {8\over 3}\left( 1 - 6\,x \right) \,\log (x) \right) ,
\end{eqnarray}
where
 ${\cal L} i_2(x)$ is the dilogarithmic function
\beq
{\cal L} i_2(x)=-\int_0^xdt {\log (1-t)\over t}.
\eeq
An important term to notice in (\ref{G2}) is the scale dependent term
proportional to $12\,G_{0}(x)\,\log( \mu^2/M_b^2 )$ in $G_2$.

The main difference between our result in Eq.~(\ref{full}) and that of
\cite{Cox1}
lies in the treatment of the large logarithmic corrections that come from the
box
diagrams, and the way they are summed using the renormalization group. In
\cite{Cox1} the $J/\Psi$ particle was treated as being a fundamental, colour
singlet field produced by the leading logarithmic effective Hamiltonian, to
which gluonic corrections were applied. This does not seem to be
correct, since
all short distance interactions that may eventually give rise to a $J/\Psi$
should be added coherently. There could possibly be other production
mechanisms, e.g., a $(v/c)^2$ suppressed (but not $L_0$ suppressed) soft gluon
induced fragmentation process. This does not, however, effect the calculation
of the hard subprocess rates. Only after the formation time $\sim 1/M_\Psi$
can the $J/\Psi$ particle be considered to be a fundamental, colour singlet
state. Indeed, as pointed out in \cite{braaten}, the renormalization group
summation of the leading logarithmic terms in \cite{Cox1} does not reproduce
the $O(\alpha_s^2)$  perturbative result correctly.
Already at the outset of the calculation in \cite{Cox1} cancellations are
performed between terms that in a correct treatment should be
multiplied by different combinations of Wilson coefficient functions.
It is therefore impossible to reconstruct the correct result from this
calculation.
Going the opposite way, trying to reproduce numerical values given in
\cite{Cox1} for some classes of diagrams from our calculations, we still find
some minor discrepancies.

In Fig.\,~1 (curves (a) and (b)) we show the leading and our next to
leading order result for
$\bpsi$. As can be seen, the next to leading order
corrections are very large and the result
still depends dramatically on the scale $\mu$. For large values of $\mu$ the
rate is even driven to unphysical negative values, suggesting that higher order
corrections are still anomalously large.

To understand this behaviour we look at the expansion of the leading order term
 $G_0  L_0^2(\mu^2)$ around some scale $\mu^*$ close to the zero point of $L_0$

\begin{eqnarray}
 L_0^2(\mu)G_0 & = & L_0^2(\mu^*)G_0 +
 {{ \alpha_s(\mu^*) C_F} \over { 4\pi }}
    12 G_0 L_0(\mu^*) L_2(\mu^*) \log (\mu^2 /{\mu^*}^2) + \nonumber \\
 & &
 \left({ {\alpha_s(\mu^*) C_F} \over { 4\pi }} \right)^2
    \left( 36 G_0 L_2^2(\mu^*) \log^2 (\mu^2 /{\mu^*}^2) +
  {\cal O} \left( L_0(\mu^*) \right) \right)
 \label{L0expansion} \end{eqnarray}
We note that the renormalization scale dependence is dominated by the
second order term in this expansion.
In the next to leading order (\ref{full}), the first order term in
(\ref{L0expansion}) is
cancelled by the $\mu$ dependent term in $G_2$ (\ref{G2}):

\newpage
\begin{eqnarray}
-{{ \alpha_s(\mu) C_F} \over { 4\pi }} 12G_0 L_0(\mu) L_2(\mu) \log (\mu^2
/{\mu^*}^2)
   & = & \nonumber \\
-{{ \alpha_s(\mu^*) C_F} \over { 4\pi }}
    12 G_0 L_0(\mu^*) L_2(\mu^*) \log (\mu^2 /{\mu^*}^2)
 & -  & \nonumber \\
 \left({ {\alpha_s(\mu^*) C_F} \over { 4\pi }} \right)^2
    ( 72 G_0 L_2^2(\mu^*) \log^2 (\mu^2 /{\mu^*}^2)  & + &
   {\cal O}\left(L_0 \left( \mu^* \right)\right) )
 \label{G2expansion} \end{eqnarray}
After the cancellation only a second order $\mu$ dependent term remains.
This term is twice as large as the one dominating the renormalization
scale dependence of the leading order term (\ref{L0expansion}), however, and
moreover
has the opposite sign.

To conclude, the ${L_0}^2$ suppression postpones the cancellation of the
leading order renormalization scale dependence until the second order.
The next to leading order corrections merely mirror the leading order
scale dependence. Thus, if the appropriate scale for the process is close
to the zero point of $L_0$ (curve (a) in Fig.\,~1) the qualitative picture of
the scale dependence in the next to leading logarithmic approximation
(curve (b) in Fig.\,~1)  is an inevitable consequence of the strong
scale dependence and suppression in the leading logarithmic approximation.

To overcome this problem we propose a simultaneous expansion in $L_0$ and
$\alpha_s$.
This amounts to a rather modest resummation, mixing terms from three orders in
$\alpha_s$. In this way terms proportional to $L_0$ or ${L_0}^2$ are added
together
with the terms that cancel their extreme scale dependence. Since we have
already
seen that we do not gain any information by adding these terms separately
we do not
 lose anything by doing the expansion this way.  Let us now analyze if we can
gain something.

In the new $L_0$-$\alpha_s$ expansion
we find that the old leading term ${L_0}^2 G_0$ is replaced by the square of
the two bremsstrahlung diagrams from the $c\overline{c}$ pair --- the $G_3$
term in
Eq. \,~(\ref{full}).
We note that the $G_3$ term dominates ${L_0}^2 G_0$
only at scales close to the zero point of $L_0$. From the argument above we
know, however, that the extreme scale dependence of ${L_0}^2 G_0$ will be
cancelled in
the new scheme, and it is only after this cancellation that we can expect to
get something
numerically small, as compared to the $G_3$ term.
The correct question to ask is obviously - will the next to leading order term
in the $L_0$-$\alpha_s$  expansion be numerically small?

In a complete next to leading order calculation in the new scheme one is forced
to
calculate all ${\alpha_s}^2$ terms that are not $L_0$ suppressed. Such a
calculation
could possibly be performed since all corrections coming from two-loop diagrams
or
from higher order matching and evolution of the effective weak theory are $L_0$
suppressed.

We have already calculated several of the terms needed in such a new next to
leading order calculation.
Lacking a complete calculation we try to get as much information as possible
from
the calculated terms and from the structure of the unknown terms.
The missing terms are all proportional to ${\alpha_s}^2 {L_2}^2$ and are
therefore
$\alpha_s$ suppressed as compared to the leading order term. We know, however,
that
they contain the term $36 G_0 \log^2(\mu^2)$ that cancels the scale dependences
in the $G_0$ and $G_2$ terms (\ref{L0expansion},\ref{G2expansion}). It is the
large numerical prefactor of this $\log^2$ term that threatens the viability
of the expansion.

The $G_2$ term in eq. (\ref{full}) is dominated by the cross term between a
group
of one-loop diagrams and the born graph. It is the square of this group of
diagrams that contain the dangerous $\log^2$ term.
Since the UV-divergences of the one loop diagrams have the same gamma matrix
structure
as the born diagram, we can use ${G_2}^2/(4G_0)$ as a rough estimate of the
square
of the one-loop diagrams. If we add this approximate term to the next to
leading order
terms that we have calculated exactly, we get

\begin{eqnarray}
 & & \Gamma_{_{ B \rightarrow J/\psi + X }}\approx  {\alpha_s(\mu) \over 4\pi}
C_F G_3 {L_2}^2 \times
                     \nonumber \\ & &
   \left(1 - {\alpha_s(\mu) \over 4\pi}C_F{2B \over N} +
   {{  \Delta\alpha_s } \over {4\pi}} C_F
     \left( {{2R_{+}}\over{N+1}} + {{2R_{-}}\over{N-1}} \right) \right) +
                     \nonumber \\ & &
 G_0\left( {L_0}^2 +
    {{  \Delta\alpha_s  } \over {4\pi}} C_F
      L_{0}L_{2} 2R_{1}  +
    \left({{ \Delta\alpha_s} \over {4\pi}} \right)^2 C_F^2
      {L_2}^2 {R_1}^2 \right) -
                     \nonumber \\ & &
 {\alpha_s(\mu) \over 4\pi} C_F\left( G_{2} - 2B G_{0} \right) \left( L_0 L_2 +
    {{  \Delta\alpha_s } \over {4\pi}} C_F
      {L_2}^2 R_{1} \right) +
                     \nonumber \\ & &
 \left({\alpha_s(\mu) \over 4\pi}\right)^2 C_F^2 {L_2}^2 {\left( G_{2} - 2B
G_{0} \right)^2 \over 4G_0}
 , \label{newNLO}
\end{eqnarray}
where $R_1 \equiv R_{+}-R_{-}$ and
$\Delta\alpha_s \equiv   (\alpha_{s}( M_W) -\alpha_s(\mu))$.

Numerically we find that the cancellation is quite
effective (Fig.\,~1 (c)). Remember that the born term, which is several times
larger than
the new leading order term at the $M_b$ scale, is one of the terms in the
correction.
This incomplete second order calculation should not be viewed as an
approximation
to the complete one. It is only meant to show that the cancellation between
the terms which are the largest, viewed  separately, does take place.
We have completely neglected the double bremsstrahlung and one-loop
bremsstrahlung
diagrams appearing at the ${\alpha_s}^2 {L_2}^2$ level.
The next to leading order correction in the $L_0$-$\alpha_s$  expansion could
very well be substantial, but
there is reason to believe that it is at least not larger than the leading
term.
In Fig.\,~2 we display the scale dependence of the relevant ingredients
in our scheme $L_0$, $L_2$ and $\alpha_s$ separately.

In the numerical calculations we used $\Lambda_{QCD}^{(5)}=190$ MeV
($\alpha_s(M_Z)=0.115$),
$M_b=5$ GeV and  $x\equiv 4M_c^2/M_b^2 =(M_\psi/M_{B^0})^2
\approx 0.38$.

Varying $\Lambda_{QCD}$ and $M_b$ (with $M_c/M_b$ fixed) we find - not
surprisingly - that the
$L_0$-$\alpha_s$ expansion
works best for large $\alpha_s$ and low $M_b$ when the zero point of $L_0$ is
close to the
$M_b$ scale. With $\alpha_s(M_z)=0.120$ the cancellation is almost perfect
while it is
already beginning to break down at $\alpha_s(M_z)=0.110$.
Furthermore, we note that the hard gluon emission term $G_3$ is more sensitive
to variations
in $M_c/M_b$ or $M_s/M_b$ (see \cite{Per} for expressions to first order in
$M_s^2/M_b^2$)
than the other terms. Thus a 5\% lower value for $M_c/M_b$ results in a 20\%
increase in the
leading order term in the $L_0$-$\alpha_s$  expansion. If we increase $M_s/M_b$
from 0 to 0.1
corresponding to $M_s \approx 0.5$ GeV we find a 5\% suppression of $G_0$ while
$G_3$ is
suppressed by substantial 17\%.

We have calculated next to leading order corrections also for $\eta_c$
production in B-meson decays,
and find an almost identical pattern as in the case of $J/\psi$
production. To check the scheme dependence of our calculations, we have
also utilized the naive dimensional regularization (NDR) scheme, with
negligible differences in the final results.

To be able to compare our results with experiments we first have to deal with
the strong
overall dependence on heavy quark masses, the Kobayashi-Maskawa matrix element
$V_{cb}$ and
on  $R_S(0)$, the value of the radial wave function for the $c\overline{c}$
pair
at zero distance.
Using the measured rate ($5.36 \pm 0.29$ keV) for the electromagnetic decay of
the $J/\psi$
\cite{ParticleData92} and the next to leading order result
\beq
 \Gamma_{_{ J/\psi \rightarrow e^+e^-  }}  =
         {{16\alpha_E^2 \left|R_S(0)\right|^2}\over {9{M_\psi}^2}}
        \left( 1-{16\alpha_s\over {3\pi}} \right) \label{emdecay}
\eeq
we find $\left| R_S(0) \right|^2 / {M_c}^2 \approx 0.5$ GeV. This must,
however,
be considered a very rough estimate since the QCD corrections in
(\ref{emdecay}) are
extremely large. In
the leading order we would get $\left| R_S(0) \right|^2 / {M_c}^2 \approx 0.2$
GeV. Hopefully
lattice calculations will soon give more precise estimates.
To cancel the $V_{cb}$ dependence and to reduce the very strong $M_b$
dependence
we normalize our results to the semileptonic branching ratio.
To next to leading order in $\alpha_s$ the semileptonic decay rate is
\cite{altarelli}
\beq
 \Gamma_{_{ SL }}  =
    {{{G_{F}}^2}\,{\left|{V_{cb}}\right|^2}
    \over {192\,{{\pi }^3}}}
    {M_{b}}^5 g(M_c/M_b) \left( 1 - {{2 \alpha_s(\mu)} \over {3 \pi}}
f(M_c/M_b) \right)
  \label{semil}
\eeq
where
\beq
 g(x) = 1 - 8x^2 -24x^4\log(x) + 8x^6 - x^8
\eeq
and $f(x)$ can be found in tabulated form in \cite{cabibbo}.
Using $ M_c/M_b = M_{D^0}/M_{B^0} \approx 0.35$  in (\ref{semil}) with
$\alpha_s$ at 5 GeV
and the measured semileptonic branching ratio ($10.7 \pm 0.5$)\%
\cite{ParticleData92} we find
\beqa
   Br_{_{ B \rightarrow J/\psi + X }}
 & \equiv & {{\Gamma_{_{ B \rightarrow J/\psi + X }}}\over {\Gamma_{tot}}}
             = \left({{ Br_{_{ SL }}  G_0} \over
                { \Gamma_{_{ SL }}}} \right)
        {{ \Gamma_{_{ B \rightarrow J/\psi + X }}} \over {G_0}} \nonumber \\
     & \approx &  (0.2 \%)\times \left( {5 GeV \over M_b}\right)
                      \left( {{\left| R_S(0) \right|^2 / {M_c}^2} \over {0.5
GeV}} \right)
                \left({{ \Gamma_{_{ B \rightarrow J/\psi + X }}/G_0}\over
0.05}\right)
\eeqa
The measured inclusive branching fraction for $\psi$ production in B decays is
$(1.12\pm 0.16)$\% . Using measured branching fractions
for B decays into $\psi'$, $\chi_{c1}$ and for the cascade decays
$\psi'\rightarrow \psi + X$, $\psi'\rightarrow \chi_{c1} + \gamma$ and
$\chi_{c1}\rightarrow \psi + \gamma$, one finds the branching fraction
$(0.71\pm 0.20)$
\cite{braaten} for direct $J/\psi$ production in B decays, assuming that no
other
cascade decays give significant contributions to the $J/\psi$ production.
Using the large value 0.5 GeV for $\left| R_S(0) \right|^2 / {M_c}^2$ and
taking the
size of the QCD-suppression to be $\sim 0.05$ from Fig.\,~1, we are still two
and a half
standard deviations from the experimental result.

We thus conclude from
our analysis that unless the experimental value goes down
as more data accumulate, it seems that the decay $B\to J/\Psi + X$ is a
process that cannot presently be well described by a standard application of
perturbative QCD to zeroth order in the relative velocity of the
$c\overline{c}$
quarks. We have pointed out that the strong QCD suppression of the colour
singlet operator give rise to strong cancellations between terms appearing
in different orders in $\alpha_s$.
Although we have suggested a method to ameliorate this problem, there is room
for other QCD effects, subleading in most other cases, that could bridge the
gap
between the predicted and observed values for the branching ratio. It is
thus not at all necessary at this point to make the conclusion that exotic
mechanisms (i.e. beyond the Standard Model) are needed to explain the large
experimental rate for this type of decay.

We thank S. {\AA}minneborg and R. Robinett for useful discussions. This
work was supported by the Swedish Natural Science Research Council.

\newpage
\vskip .2cm
\begin{flushleft}
\parindent 0pt {\Large Figure Captions}
\end{flushleft}
\vskip .3cm
\begin{itemize}
\item[Fig.\,~1.]
Renormalization scale dependence of $\Gamma_{B\to J/\Psi + X}$ normalized
to the naive parton model result $G_0$, in various approximations:
\begin{itemize}
\item[(a)] Leading logarithmic approximation (for consistency, the leading
order result for $\alpha_s$ is used in the definition of $L_\pm$).
\item[(b)] Conventional next to leading logarithmic approximation.
\item[(c)] Leading order result in the proposed $L_0$-$\alpha_s$ expansion.
\item[(d)] Incomplete next to leading order result in the
$L_0$-$\alpha_s$ expansion, as given by Eq.\,~(\ref{newNLO}).
\end{itemize}
\item[Fig.\,~2.] Renormalization scale dependence of the colour singlet and
octet Wilson coefficient functions $L_0$, $L_2$ and of $\alpha_s$.

\end{itemize}


\newpage
\begin{thebibliography}{99}

\bibitem{braaten} G. Bodwin, E. Braaten T.C. Yuan and G.P. Lepage,
Phys. Rev. {\bf D46} (1992) 3703.

\bibitem{altarelli} G. Altarelli and S. Petrarca,
\newblock  Phys. Lett. {\bf {B261}} (1991) 303.

\bibitem{buras} A. Buras and  P.H. Weisz, Nucl. Phys. {\bf B333}
(1990) 66.

\bibitem{bigi} I. Bigi, B. Blok, M.A. Shifman and A. Vainshtein, CERN-
TH.7082/93, hep-ph 9311339, 1993.

\bibitem{Cox1} P.W. Cox, S. Hovater, S.T. Jones and L. Clavelli,
Phys. Rev. {\bf D32} (1985) 1157.
\bibitem{Per} P. Ernstr\"om, Stockholm University report and PhD thesis, in
preparation.
\bibitem{gaillard} M.K. Gaillard and B.W. Lee, Phys. Rev. Lett. {\bf 33}
 (1974)  108.
\bibitem{maiani} G. Altarelli and L. Maiani, Phys. Lett. {\bf 52B}  (1974)
351.
\bibitem{kuhn} M.B. Wise, Phys. Lett. {\bf 89B} (1980) 229;
T.A. DeGrand and D. Toussaint, Phys. Lett. {\bf 89B} (1980) 256;
J.H. K\"uhn, S. Nussinov and R. R\"uckl, Z. Phys. {\bf C5} (1980) 117.
\bibitem{HV} G. 't Hooft and M. Veltman, Nucl. Phys. {\bf B44} (1972) 189.
\bibitem{grinstein} B. Grinstein, Phys. Lett. {\bf B229} (1989) 280.
\bibitem{buras2} A.J. Buras, M. Jamin, M.E. Lautenbacher and P.H. Weisz,
Nucl. Phys. {\bf B370} (1992) 69.

\bibitem{ParticleData92} K. Hikasa et al. (Particle Data Group),
Phys. Rev. {\bf D45}, Part 2 (1992) 3703.

\bibitem{cabibbo} N. Cabibbo and L. Maiani,
\newblock  Phys. Lett. {\bf {B79}} (1978) 109.
\end{thebibliography}
\end{document}